\documentclass{revtex4}
\usepackage{graphicx}
\usepackage{amsmath}
\usepackage{amssymb} 

\begin{document}

\title{Ferrell-Berreman modes in plasmonic epsilon-near-zero media}
\author{Ward D. Newman}
\affiliation{Department of Electrical and Computer Engineering, University of Alberta, Edmonton AB, Canada, T6G 2V4}
\author{Cristian L. Cortes}
\affiliation{Department of Electrical and Computer Engineering, University of Alberta, Edmonton AB, Canada, T6G 2V4}
\author{Jon Atkinson}
\affiliation{Department of Electrical and Computer Engineering, University of Alberta, Edmonton AB, Canada, T6G 2V4}
\author{Sandipan Pramanik}
\affiliation{Department of Electrical and Computer Engineering, University of Alberta, Edmonton AB, Canada, T6G 2V4}
\author{Raymond G. DeCorby}
\affiliation{Department of Electrical and Computer Engineering, University of Alberta, Edmonton AB, Canada, T6G 2V4}
\author{Zubin Jacob}
\email{zjacob@ualberta}
\affiliation{Department of Electrical and Computer Engineering, University of Alberta, Edmonton AB, Canada, T6G 2V4}

\begin{abstract}
We observe unique absorption resonances in silver/silica multilayer-based epsilon-near-zero (ENZ) metamaterials that are related to radiative bulk plasmon-polariton states of thin-films originally studied by Ferrell (1958) and Berreman (1963).
In the local effective medium, metamaterial description, the unique effect of the excitation of these microscopic modes is counterintuitive and captured within the complex propagation constant, not the effective dielectric permittivities. Theoretical analysis of the band structure for our metamaterials shows the existence of multiple Ferrel-Berreman branches with slow light characteristics. The demonstration that the propagation constant reveals subtle microscopic resonances can lead to the design of devices where Ferrell-Berreman modes can be exploited for practical applications ranging from plasmonic sensing to imaging and absorption enhancement. 
\end{abstract}

\maketitle



An important class of artificial media are the epsilon-near-zero (ENZ) metamaterials that are designed to have a vanishing dielectric permittivity $|\epsilon|\rightarrow0$.
Waves propagating within ENZ media have a divergent phase velocity that can be used to guide light with zero phase advancement through sharp bends within sub-wavelength size channels \cite{Huang2011,Silveirinha2006}, or to tailor the phase of radiation/luminescence within a prescribed ENZ structure \cite{Alu2009,Vesseur2013}. The electric field intensity within an ENZ medium can be enhanced relative to that in free space leading to strong light absorption \cite{Kats2012}. This enhanced absorption in ENZ media has been exploited for novel polarization control and filtering in thin films \cite{Alekseyev2010}, as well the proposal to use ENZ absorption resonances to tune thermal blackbody radiation of a heated object to the band-gap of a photovoltaic cell \cite{Molesky2013}. An enhanced non-linear response based upon  strong spatial dispersion of waves in ENZ media has been demonstrated, and proposed for all-optical switching \cite{Pollard2009,Wurtz2011}.

Here we show theoretically and experimentally that ENZ metamaterials support unique absorption resonances related to radiative bulk plasmon-polaritons of thin metal films. 
These radiative bright modes exhibit properties in stark contrast to conventional dark modes of thin-film media (surface plasmon polaritons).
The unique absorption resonances manifested in our metamaterials were originally studied by Ferrell in 1958 for plasmon-polaritonic thin-films in the ultraviolet \cite{Ferrell1958}, and by Berreman in 1963 for phonon-polaritonic thin-films in the mid-infrared spectral region \cite{Berreman1963}.
Surprisingly, two research communities have developed this independently with little communication or overlap until now: we therefore address these resonances as Ferrell-Berreman (FB) modes of our metamaterials.
Counterintiutively, in the metamaterial effective medium picture, these resonances are not captured in the metamaterial dielectric permittivity constants but rather in the effective propagation constant.
Furthermore, we show the existence of multiple branches of such FB modes that have slow light characteristics fundamentally different from the single thin film case.
Our work can lead to applications where FB modes are used for thin-film characterization \cite{lee2006all}, sensing \cite{liu2009planar}, imaging \cite{kubo2005femtosecond}, absorption enhancement \cite{wang2010broadband} and polarization control \cite{sukharev2006phase}.

	An isotropic ENZ occurs naturally in metals such as silver and aluminum, and in polar dielectrics such as silicon carbide and silicon dioxide.
	However, due to the low effective mass of electrons, the ENZ inevitably occurs in the ultraviolet (UV) spectral range for metals ($\omega_p \propto 1/\sqrt{m_e}$).
	On the other hand the large effective mass of ions shifts the ENZ at the longitudinal optical phonon frequency to the the infrared (IR) spectral range for polar dielectrics ($\omega_{LO} \propto 1/\sqrt{M_{ion}}$).
	Thus very few natural materials exhibit ENZ behaviour in the optical frequency range, and designed artifical media must be used for ENZ-based applications in the visible \cite{Wurtz2011,Maas2013}.
	Here, we design ENZ media using silver (Ag) /silica SiO$_2$ multilayers with nanoscale, sub-wavelength layer thicknesses to achieve a tunable anisotropic ENZ at optical frequencies.

	 
	Zeroth order Maxwell-Garnett Effective Medium Theory (EMT) shows that the permittivity of multilayer structures composed of a subwavelength thickness metal/dielectric unit cells with dielectric constants $\epsilon_m$ and $\epsilon_d$, and layer thicknesses $d_m$ and $d_d$ is indeed uniaxially anisotropic.
	The response of the multilayer is described via a dielectric permittivity tensor of the form $\bar{\bar{\epsilon}}_{\text{eff}} = \textrm{diag}[\epsilon_{||},\epsilon_{||},\epsilon_\perp]$, where
	$\epsilon_{||} = \rho\epsilon_{m} + (1-\rho)\epsilon_{d}$ is the permittivity for polarizations along the layer interfaces, and $\epsilon_\perp = \left(\rho/\epsilon_{m}+(1-\rho)/\epsilon_{d}\right)^{-1}$ is the permittivity for polarizations perpendicular to the layer interfaces.
	$\rho = d_m/(d_d+d_m)$ is the metal volume filling fraction.

In this paper we apply the local EMT model that is valid for free space wavelengths that are much longer than the multilayer unit cell thickness equal to $d_m+d_d$. However, it has been shown that this local EMT model fails to accu-rately describe the highly confined Bloch surface plasmon-polariton modes of metal/dielectric superlattices (so called high-$k$ waveguide modes of hyperbolic metamaterials). There has been significant development of a non-local EMT that more accurately describes the high-$k$ waveguide modes of metal/dielectric superlattice-based metamaterials \cite{Orlov2013,Orlov2011,Chebykin2010,Chebykin2011}.	In this paper we study photonic modes that can be excited from free space (exist within the light cone) and are therefore not deeply sub-wavelength. Our simulations of practical multilayer structures beyond the effective medium model consider the role of finite unit cell size, absorption and dispersion. The results show strong agreement with experimental observations.

	The permittivity of almost all traditional metals can be approximated using free-electron Drude dispersion and using this fact, we see that the parallel dielectric constant $\epsilon_\parallel$ is a modified Drude-like dispersion, and the presence of the weakly dispersive dielectric layers effectively adds a positive background dielectric constant.
	As a result, the dielectric layers serve to dilute the metal and red-shift the effective plasma frequency of the multilayer superlattice.
	The metamaterial has an ENZ in the parallel direction at the spectral frequency satisfying the relation $\rho\epsilon_m(\omega_{ENZ})=-(1-\rho)\epsilon_d$.
 	We can thus tune the parallel permittivity ENZ frequency by choosing an appropriate metal $\epsilon_m(\omega)$, dielectric $\epsilon_d$, and fill fraction $\rho$.
	Similarly, the perpendicular permittivity $\epsilon_\perp$ has a Lorentz-like response with a resonance pole at the spectral frequency satisfying the relation $(1-\rho)\epsilon_m(\omega_{LP})=-\rho\epsilon_d$.
	The Lorentz-resonance pole for the perpendicular permittivity $\epsilon_\perp$ is blue-shifed to higher frequencies with decreasing fill fractions of metal while the resonance pole quality is reduced.

\begin{figure} 
		\centering
		\includegraphics{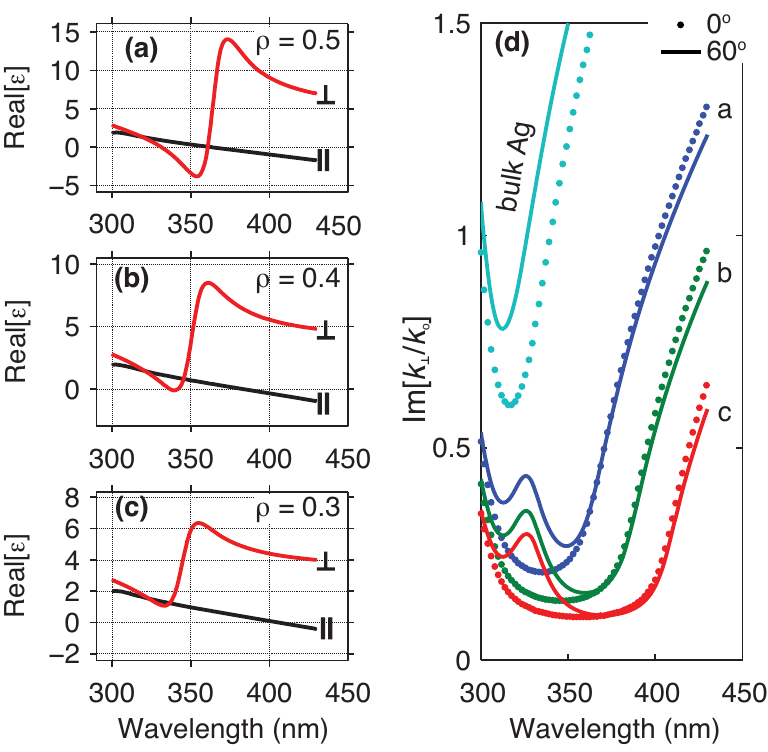}
		\caption{The real part of the dielectric permittivity tensor is shown for silver/silica multilayer metamaterials with a  silver thickness of 20 nm and silica thicknesses (a) 20 nm, (b) 30 nm, and (c) 40 nm. The dispersion is calculated using experimentally obtained dielectric permittivities for the constituents. (d) The imaginary part of the complex propagation constant $k_\perp$ for $p$-polarized light is shown for the metamaterials samples in panels (a)-(c), and for bulk silver. $\textrm{Im}[k_\perp^p]$ governs the transparency window and exhibits an anomalous peak at the ENZ of of the constituent silver ($\lambda\approx326$ nm) for obliquely incident light.}
		\label{EMTDispAndkz}
\end{figure}

	{\section{Experimental Evidence of Radiative Bulk Plasmons in ENZ Metamaterials}}
	Figure \ref{EMTDispAndkz}(a)-(c) shows the calculated dispersion of the effective medium dielectric constants for silver/silica metamaterials with various fill fractions of silver. 
	The fill fractions shown in Figure \ref{EMTDispAndkz} correspond to multilayers with silver thickness 20 nm and silica thickness 20, 30 and 40 nm.
	We see that the ENZ in the parallel direction can be spectrally tuned by varying the relative fill fraction of the silver within the metamaterial. We also note that a resonant ENZ effect can occur in the perpendicular component of the dielectric constant as well; however the dissipative, lossy component of the perpendicular permittivity is relatively large with Im$[\epsilon_\perp]\sim1-10$ in the perpendicular direction. This severely weakens the observable ENZ effects. Furthermore, we emphasize that for small enough fill fractions of silver, the Lorentz-resonance pole in the permittivity in the perpendicular direction is weakened to that point where the real part of the perpendicular permittivity does not vanish at all, as shown in Figure \ref{EMTDispAndkz}(c).
	
	We now discuss wave propagation through multilayer metamaterial slabs.
	The dispersion of $p$-polarized plane waves with wave vector $\vec{k}$ and spectral frequency $\omega$ is described by the equation $k_{\parallel}^2/\epsilon_\perp+k_\perp^2/\epsilon_\parallel=k_o^2$ for waves in the metamaterial. 
	We define $k_\perp = \sqrt{\epsilon_\parallel(k_o^2-k_\parallel^2/\epsilon_\perp)}$ as the propagation constant of waves in the metamaterial.
	Figure \ref{EMTDispAndkz}(d) shows the imaginary part of the propagation constant for the three metamaterial samples shown in Figure \ref{EMTDispAndkz}(a)-(c),and for bulk silver.
	We see the metamaterials possess a spectral range where the attenuation of waves within the medium is low (small Im$[k_\perp]$).
	Over this spectral range the metamaterial is effectively dielectric with Re$[\epsilon_{\parallel,\perp}]>0$.
	The real part of the propagation constant in the metamaterial Re$[k_\perp]$ is close to the propagation constant of vacuum and $\epsilon_{\parallel}\approx\epsilon_o=1$, thus resulting in a low reflectivity.
	This spectral range where both Im$[k_\perp]$ and reflectivity are small defines the transparency window of the metamaterial.
	Transparency windows have been exploited for optical filters in the UV \cite{Bloemer1998}. 

		\begin{figure*}
		\begin{center}
			\includegraphics[width=440pt]{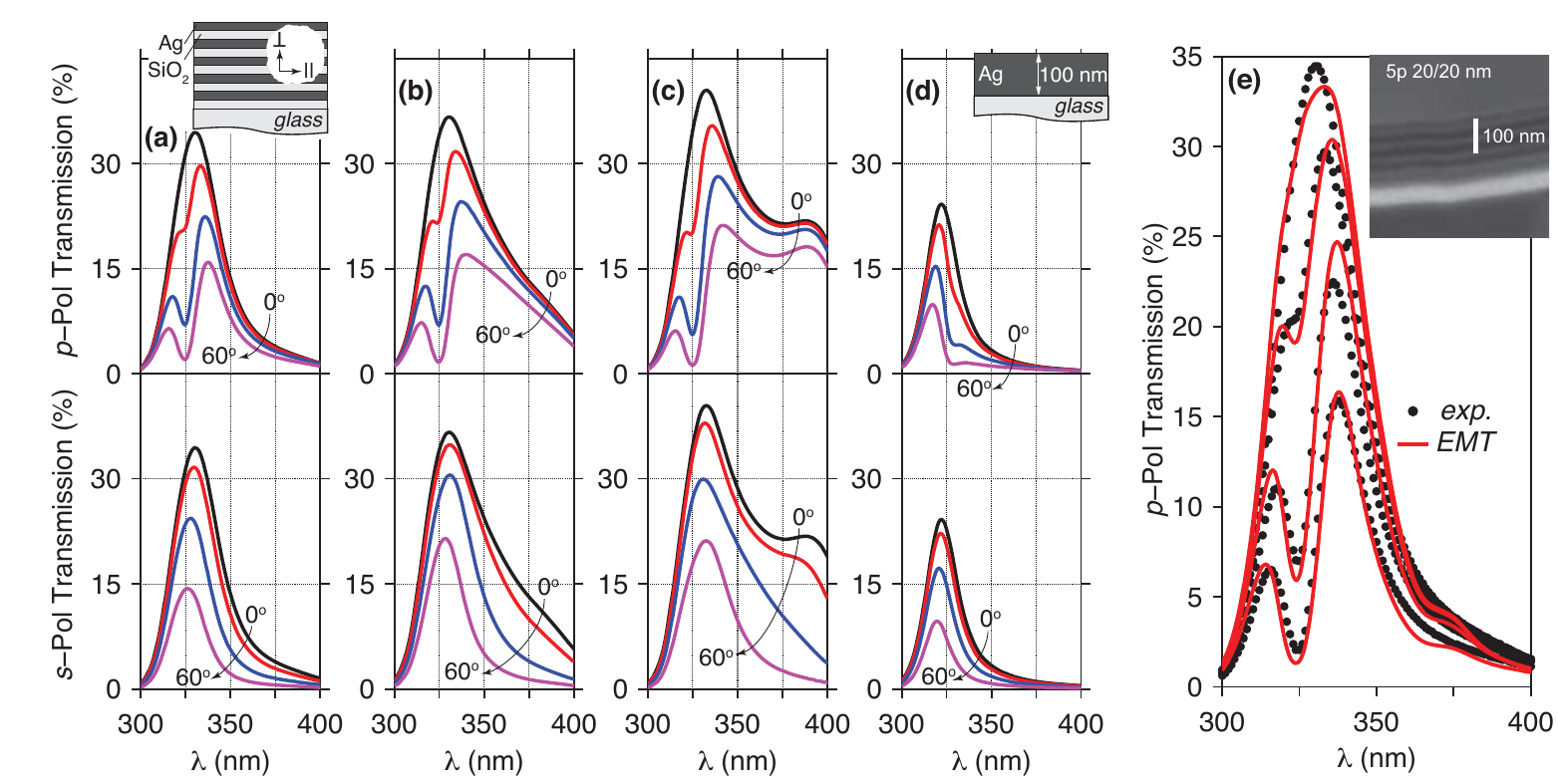}
			\caption{Experimental $s$- and $p$-polarized transmission at angles of incidence 0$^o$, 20$^o$, 40$^o$ and 60$^o$ is shown for five period silver/silica multilayer metamaterials with layer thicknesses (a) 20/20 nm, (b) 20/30 nm, (c) 20/40 nm, and for a (d) 100 nm silver film. The excitation and subsequent dissipation of Ferrell-Berreman (FB) modes within the silver films appears as anomalous transmission for obliquely incident $p$-polarized light at the ENZ frequency of the constituent silver films ($\lambda\approx 326$ nm). FB modes can not be excited with $s$-polarized light and thus no anomalous transmission is present. The FB mode is only weakly present in the relatively thick silver film. Panel (e) shows the strong agreement of EMT calculations with experiment for the five period 20/20 nm sample. {\it Inset:} SEM micrograph of the sample.}
			\label{TpTsFullPanels}
		\end{center}
	\end{figure*}

	We note that Figure \ref{EMTDispAndkz}(d) shows a counterintuitive local absorptive peak in the propagation constant for {\it p}-polarized waves obliquely incident at $\lambda \approx 326~\text{nm}$, the ENZ of silver.
	The spectral location of this peak is fixed, independent of the silver filling factor $\rho$ and is not observed in bulk silver or bulk silica.
	It should also be stressed that this absorption peak does not occur for $s$-polarized waves propagating in the metamaterials. 
	Therefore, its physical origin is not solely material absorption within the constituent layers of the metamaterial but must be due to special modal properties.
	We emphasize that the anomalous peak at $\lambda=326$ nm occurs at the ENZ of the constituent silver films and does not occur at the ENZ nor at the Lorentz-pole of the metamaterial permittivity tensor components; there is no peak in the imaginary, dissipative part of the permittivity tensor components \footnote{See supplementary information for the imaginary part of the dielectric permittivity tensors}.

	Our main aim is to show theoretically and experimentally that the physical origin of this anomalous absorption peak in metamaterial propagation constant is due to the excitation of microscopic resonances of the metamaterial: radiative bulk polaritons of thin-films which we call Ferrell-Berreman modes. 
	This absorption peak exists within the light cone of vacuum and can therefore be observed in the free-space transmission spectrum of the metamaterials.

	 We deposited five-period silver/silica multilayer metamaterials on glass microscope slides via electron beam evaporation.
	 The permittivities of silver and silica were extracted through ellipsometry on individual silver and silica films.
	 Atomic Force Microscope measurements indicate an RMS surface roughness of $\approx1-2.5\text{ nm}$.
	 For a control sample, a 100 nm silver film was also fabricated.
	 By volume, the control sample contains an equal amount of silver as the metamaterials.
	 In Figure \ref{EMTDispAndkz} we showed the extracted permittivity tensors and propagation constants, calculated using the extracted dielectric constants for the fabricated samples.
	
	Figures \ref{TpTsFullPanels}(a)-(c) show the experimentally observed {\it s}- and {\it p}-polarized specular transmission through the three silver/silica multilayer samples and through the control sample.
	Each metamaterial displays a transparency window whose width increases as the metal is diluted further (decreasing $\rho$).
	As expected from EMT predictions, the three metamaterial samples exhibit a counterintuitive $p$-polarized transmission dip at the silver ENZ $\lambda \approx 326$ nm.
	Furthermore, we observe that this anomalous dip does not depend on the silver filling fraction nor on the total metamaterial thickness.
	This clearly implies that the effect is not due to cavity Fabry-Perot resonances. 
	Another important aspect is that this anomalous transmission dip is hardly distinguishable in the 100 nm thick silver control sample (Figure \ref{TpTsFullPanels}(d) while the dip's spectral energy is slightly red-shifted from the ENZ of silver.
	Panel (e) of Figure \ref{TpTsFullPanels} shows the excellent agreement between the local EMT predictions and the experimentally observed transmission.

\section{Modal Analysis of Radiative Bulk Plasmons - Ferrell-Berreman Modes}

	Through modal analysis, taking into account the finite unit cell size of the metamaterial, we now show that the physical origin of this bulk absorption in the metamaterials is due to the excitation of leaky bulk polaritons called Ferrell-Berreman modes \cite{Vassant2012,Vassant2012a}.
	Bulk metal supports volume charge oscillations at the ENZ of the metal (bulk or volume plasmons).
	These excitations are a completely longitudinal wave and therefore can not be excited with free space light, a transverse wave.
	For films of metal with thicknesses less than the metal skin depth, the top and bottom interface couple, allowing for collective charge oscillations across the film.
	The bulk plasmon then is no longer purely longitudinal and can interact with free space light at frequencies near the metal ENZ \cite{Kliewer1967}.
	This was originally pointed out by  Ferrell for metallic foils \cite{Ferrell1958,McAlister1963,Brambring1965}, and by Berreman for polar dielectric films \cite{Berreman1963}.
	Our multilayer metamaterials support several of these radiative excitations which we call FB modes.
	
\begin{figure}
		\centering
		\includegraphics[width=240pt]{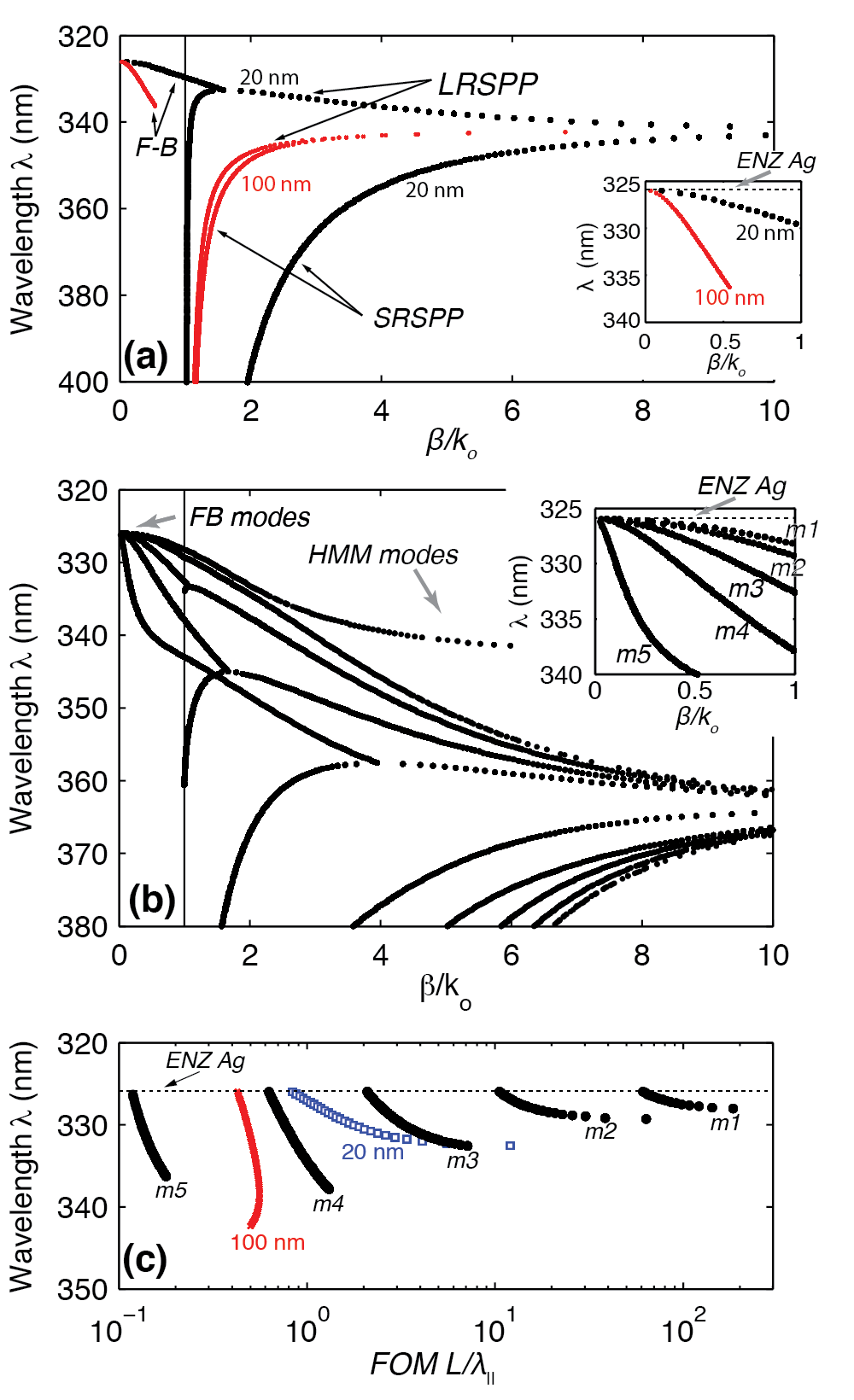}

		\caption{Ferrell-Berreman (FB) mode dispersion for (a) a 20 nm and a 100 nm thick Ag foil on glass, and (b) for a 5 period 20/30 nm Ag/SiO$_2$ multilayer on glass. The silver and silica glass is modeled in the limit of neglible material losses. (c) Figure of merit, ratio of the decay length to the wavelength, is shown for the 20 nm and 100 nm Ag foils and the 20/30 nm Ag/SiO$_2$ multilayer. The large figure of merit for the metamaterial FB modes allow them to couple strongly to free-space light. Their low group-velocity, slow-light nature leads to dissipation due ohmic losses in the metal, not re-radiation to be the dominant decay channel for the FB excitations.}
		\label{BandStructure}
	\end{figure}

	These FB modes differ from the well known surface plasmon polaritons supported by metal foils, by the fact that in surface plasmon modes, energy propagates along the surfaces of the metal, whereas in FB modes volume charge oscillations are setup across the foil and energy propagates within the bulk of the metal.
	Additionally, surface plasmon modes lie to the right of the light line and do not interact with free space light.
	The thin-film bulk polaritons we observe have transverse wavevectors similar to free space light and exist to the left of the light line.
	%
In Figure \ref{BandStructure}(a), we show the dispersion of the radiative bulk plasmon and the surface plasmons for a thin and a thick silver foil treated in the low loss limit.
The thin vertical line is the light line of vacuum.
We determine the modal dispersion by locating the poles of the reflection coefficient of the structure $r_p(k_\parallel,\lambda)$ (see supplementary info and references therein for details)
Here $k_\parallel =\beta+i\kappa$ is the complex wavevector along the direction parallel to the interfaces.
$\beta$ describes the wavelength parallel to the interface and thus phase advancement of the guided wave, while $\kappa$ describes the propagative decay or attenuation.
To the right of the light line we see the two well known surface plasmon modes of a single metallic thin-film: the long- and short-ranged surface plasmon polaritons (LRSPPs and SRSPPs).
To the left of the light line we see the radiative bulk plasmon, FB mode of the silver film which exists at energies near the ENZ of silver.
The radiative bulk plasmon mode exhibits a nearly flat anomalous dispersion with a negative group velocity $v_g = \partial\omega/\partial\beta<0$.
The negative group velocity indicates a strong presence of electromagnetic energy flow within the metal and this energy flow is opposite to the direction of phase front advancement \cite{Tournois1997}.

Even in the low loss limit, the FB modes are described with a complex propagation wavevector $k_\parallel$.
This fact immediately implies that FB modes attenuate as they propagate along the film due to loss of energy from radiation into free-space light.
On the other hand, in the low-loss limit, the surface plasmons are described by a completely real propagation wavevector $k_\parallel$ and do not radiate.

To understand the role of the multiple FB branches in thin film metamaterials, we define the figure-of-merit for the radiative FB modes to be $FOM=L/\lambda_\parallel$ where $L=1/\kappa$ and $\lambda_\parallel = 2\pi/\beta$ are the decay length and effective wavelength of the mode as it propagates along the film.
	The $FOM$ defined here is analogous to the quality factor of the excitation. 
	If $FOM  \lesssim 1$ the FB excitation is overdamped and essentially attenuates before propagating one wavelength.
	The figure-of-merit gives insight to the coupling of the FB modes with free-space radiation.

	Figure \ref{BandStructure}(c) shows the $FOM$ for the FB modes for the thin and thick silver films.
	We see that for thicker films of silver, the FOM is strongly reduced relative to thin-film silver and the radiative surface mode of thick silver films interact poorly with free space light despite lying to the left of the light-line.
	It is worth emphasizing that as the thickness of the silver film increases, the $FOM$ of the FB mode decreases, and in the limit of extremely thick silver film the figure-of-merit vanishes $FOM\rightarrow0$. Thus silver films with thicknesses greater than about four to five skin-depths of silver do not support a FB mode and optical measurements will not reveal an anomalous transmission for $p$-polarized light.
	The LRSPP and SRSPP are pure surface waves and their figure-of-merit diverges in the low loss limit $FOM \rightarrow \infty$.

	Note that despite treating the silver in the low-loss limit, the FB mode dispersion and FOM for the 100 nm silver film shown in Figure \ref{BandStructure} agrees with the experimentally observed behaviour of silver.
	The observed anomalous transmission for a 100 nm silver foil on glass, shown in Figure \ref{TpTsFullPanels}(d), is red shifted from the $\lambda = 326$ nm ENZ of silver as predicted by modal analysis.
	Furthermore, the experimental dip in $p$-polarized transmission is hardly discernible indicating poor free-space/FB mode coupling (low FOM).

	
	We have treated the silver films with a completely local dielectric model and predict the existence of a radiative bulk plasmon mode.
	The full non-local dielectric response gives rise to multiple absorption resonances called Tonks-Dattner resonances \cite{Lindau1971}.
	Experimental observation of these multiple bulk plasmon absorption resonances only becomes apparent for silver films thinner than about 12 nm, much thinner than the film thicknesses considered here. Our experimental results and single film / multilayer plasmonic band structure calculations are in agreement with Ferrell’s original prediction that the radiative polaritons in thin metal films occur at energies slightly below the ENZ frequency.
	This is in complete contrast to the Tonks-Dattner resonances that are observed at energies above the ENZ frequency.
	

	We now discuss the multiple radiative bulk plasmon, FB modes supported by metal/dielectric multilayer metamaterials that show fundamental differences from the thin film case.
	 Figure \ref{BandStructure}(b) shows the predicted modal dispersion of the radiative FB modes supported by a five period 20/30 nm silver/silica multilayer on glass in the limit of negligible material losses.
	 Analogous to the single interface surface-plasmons splitting into the LRSPP and SRSPP for a thin film of metal, the FB radiative surface plasmon of a single film splits into several radiative modes for the multilayer structure with the dominant modes existing only in a narrow spectral range below the ENZ of the silver film.
	 Figure \ref{BandStructure}(c) shows that the figures of merit for the multilayer radiative FB modes are much greater than unity and they are orders of magnitude higher than the equal thickness of bulk silver (100 nm), and thus the multilayer FB modes interact strongly with freespace light.
	 The dominant radiative state is loosely defined as the mode with the slowest group velocity and highest FOM (see mode labelled {\it m1} in Figure \ref{BandStructure}(b)-(c)).

	The slow-light nature of the FB modes leads to the anomalous transmission observed experimentally.
	As shown in Figure \ref{BandStructure}, the FB excitations in thin-films have a slow, negative group velocity and long lifetimes as the mode propagate along the film.
	When dissipation in silver is included, a competing decay or attenuation channel for the FB modes is present.
	Propagation losses due to dissipative ohmic heating in the metal is enhanced for slow light \cite{Monat2009}, and thus the dissipative ohmic losses are the dominant decay channel for the FB modes, not re-radiation \cite{Ferrell1962,McAlister1963}. 
	Therefore, we interpret the observed anomalous $p$-polarized transmission in the multilayer structures as the excitation and subsequent dissipation of the dominant FB mode. 
	
	\begin{figure}
		\centering
			\includegraphics[width=240pt]{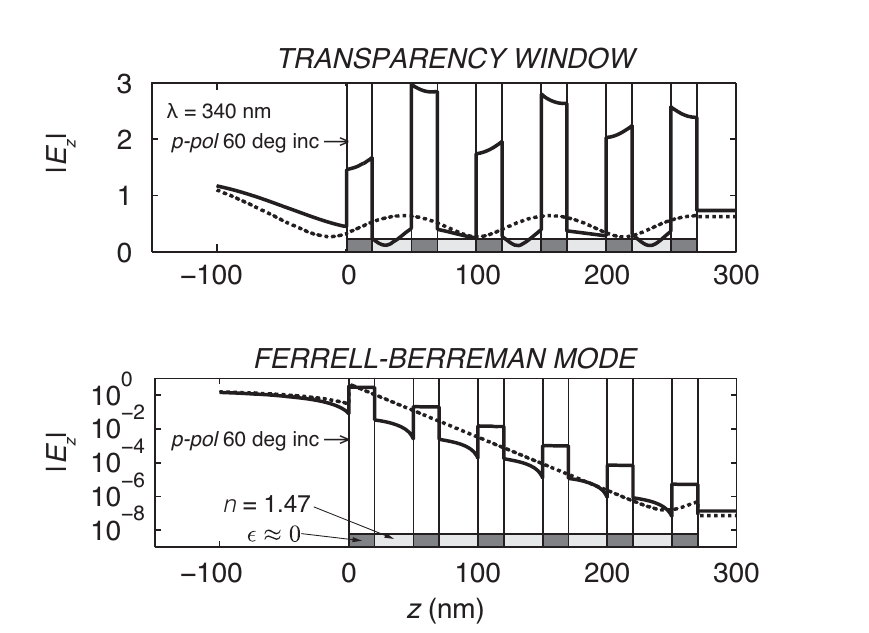}
			\caption{{\it Top: transparency window.} At $\lambda=340$ nm, {\it p}-polarized light obliquely incident on the metamaterial is transmitted without significant attenuation. {\it Bottom: Ferrell-Berreman Mode.} Within the transparency window at the plasma frequency of the constituent silver, there is an anomalous decaying mode which attenuates very rapidly and is not transmitted. Ferrell-Berreman modes are excited within the constituent silver films and charge oscillates across the volume of the silver resulting in a strong field enhancement in each silver layer. The solid curve ({\bf ---}) is the exact multilayer calculation and the dashed curve ({\bf - - -}) is the effective medium approximation. (Note the logarithmic scale in the bottom panel.)}
			\label{Fields}
	\end{figure}

	We note also that the electric field profile for the FB modes excited in our multilayer metamaterial is completely different from the well-known modes in the transparency window \cite{Bloemer1998}.
	Figure \ref{Fields} shows the predicted electric field distribution for a $p$-polarized $60^o$ obliquely incident wave upon the 20/30 nm siliver/silica multilayer.
	The top panel shows light incident at $\lambda=340$ within the transparency window where both the exact multilayer and the effective medium approximation predict that the wave is transmitted through the metamaterial without significant attenuation.
	In the bottom panel we show light incident at the ENZ of silver $\lambda\approx326$ nm.
	 The FB modes are excited within the constituent silver layers resulting in a strong attenuation of the wave (note the logarithmic scale).
	 In the exact multilayer treatment we see a strong field enhancement due to the excitation of radiative bulk plasmon modes within the metal layers.
	 Charge oscillations are setup across the volume of the silver films resulting in a static capacitor-like electric field profile.
	 In the effective medium picture however, the wave attenuates continuously.
	 We emphasize, that this bulk-like loss in the effective medium approximation is highly angle and polarization dependent.
In stark contrast, the normally incident and {\it s}-polarized waves are transmitted at the ENZ of the silver.

To summarize we have shown that metal/dielectric superlattice based epsilon-near-zero metamaterials exhibit unique micropscopic radiative bulk plasmon resonances called Ferrell-Berreman modes that can be excited with free-space light.
In the metamaterial, effective medium picture the excitation of these modes is captured in the propagation constant, not in the effective dielectric permittivity constants.
We observe these modes as anomalous transmission minima which lie within the transparency window of the metamaterials.
These radiative volume polaritonic modes could be exploited in applications such as sensing, imaging and absoprtion spectroscopy.
\\
\\

	The authors wish to acknowledge the financial support from the National Science and Engineering Research Council of Canada (NSERC),  and  from Alberta Innovates - Technology Futures.

	Supporting information available: Fabrication details, extracted optical constants, effective medium theory calculations, and modal analysis techniques are supplied in an additional supplementary information document.



\section{TOC Entry}
\begin{center}
	\includegraphics{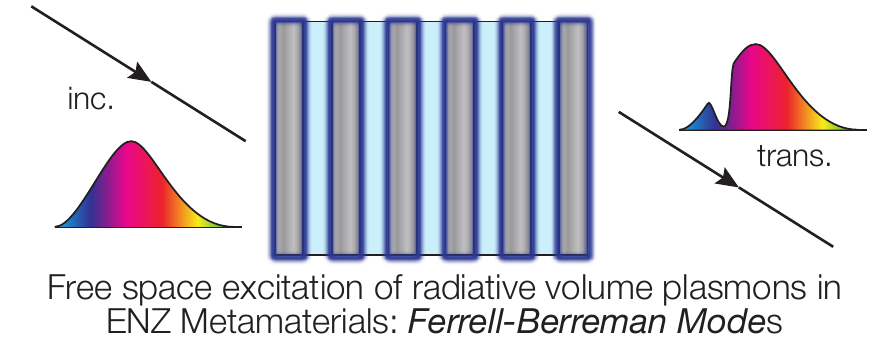}
	\end{center}

\clearpage
\section{Supplementary Info}

\subsection{Fabrication, Characterization, and Measurements}

	\subsubsection{Fabrication}
	Silver and silica multilayer films were deposited using electron beam evaporation on BK7 glass microscope slides.
	A 40 nm wetting layer of silica was deposited onto the substrate to help remove initial surface roughness.
	The silver and silica multilayers were then deposited and capped with a top 40 nm layer of silica to help prevent oxidation of the silver films.
	Film thicknesses were verified through {\tt JEOL 6301F} scanning electron microscope micrographs.
	RMS surface roughness was measured with an {\tt Alpha-Step IQ} contact profilometer and was estimated to be between 1-2 nm for all layers.

	\subsubsection{Determining Optical Constants}
\begin{figure}
\centering
	\includegraphics[width=\textwidth]{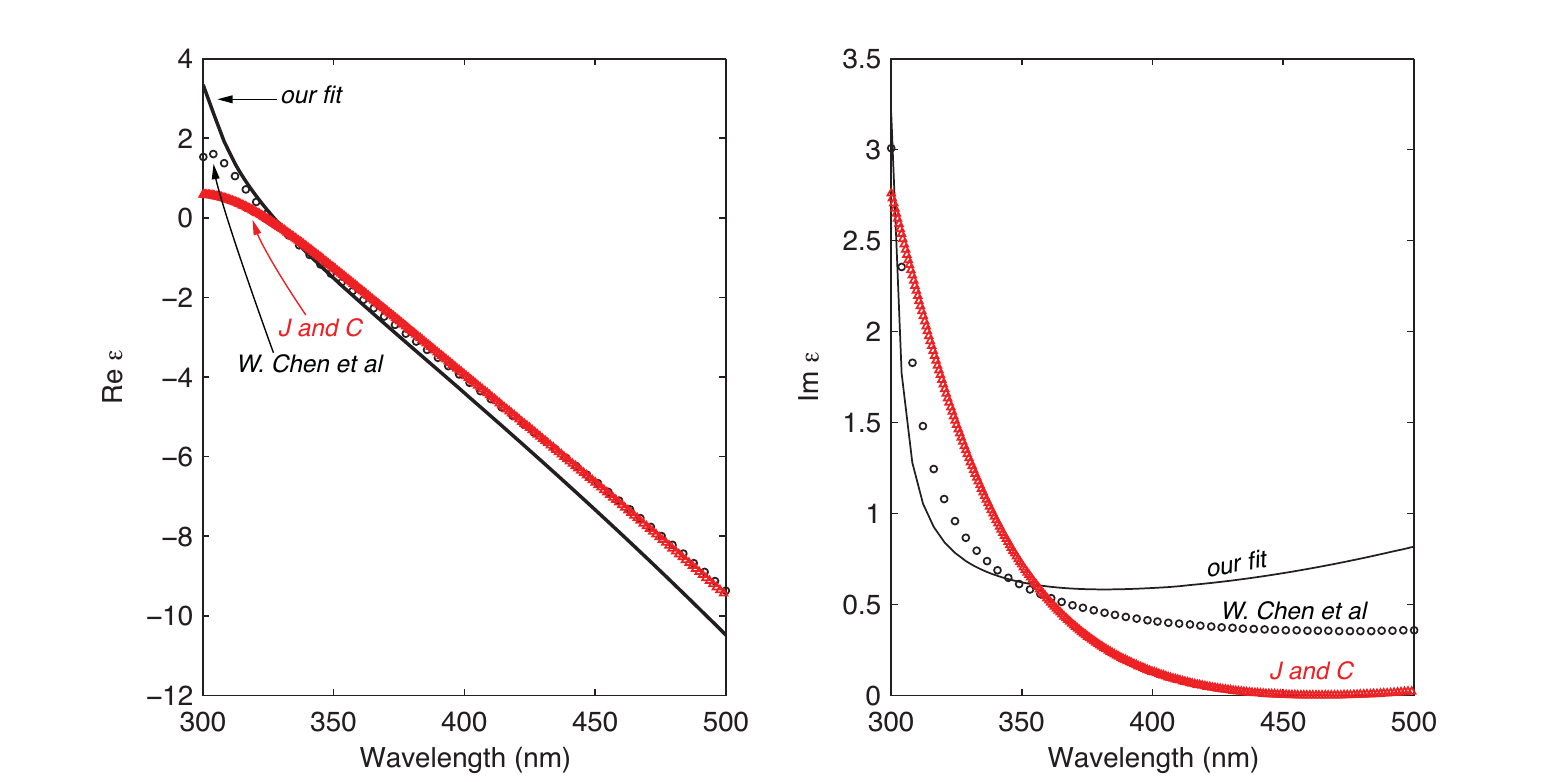}
	\caption{Dielelectric permittivity for a 100 nm thick silver foil as determined by ellipsometry.}
	\label{epsAg}
\end{figure}

	Optical constants of individual silica and silver films were determined via ellipsometry using a {\tt JA Woollam VASE} ellipsometer and the inferred optical constants showed good agreement with literature \cite{Johnson1972,Chen2010}.
	We modeled the dielectric permittivity function of silver as $\epsilon_{Ag}\approx \epsilon_{FC} + \epsilon_{bound}$, with $\epsilon_{FC}$ being the contribution from the free carriers and $\epsilon_{bound}$ being the contribution of bound or valence electrons. Free electron motion in the silver is treated with a Drude dispersion while interband transition of bound electrons in the UV are modeled with five Lorentz osscillators as described by Chen et al \cite{Chen2010,Fujiwara2007}:
	\begin{eqnarray}
		\epsilon_{fc} &\approx& \epsilon_{drude} = 1 - \frac{\omega_p^2}{\omega^2+i\omega/\tau} \\ 
		\epsilon_{bound} &\approx& \epsilon_{bg}+\sum_{n=1}^5 \epsilon_{lorentz} = \epsilon_{bg} + \sum_{n=1}^5 \frac{A_n \omega_n^2}{\omega_n^2-\omega^2-i\omega/\tau_n}.
	\end{eqnarray}
	$\omega_p$ is the plasma frequency of the metal, $\tau$ is the free electron scattering time, $\omega_n$ is the $n^{th}$ interband transition frequency, $\tau_n$ is the dissipation time of the interband excitation, and $A_n$ is the amplitude of the interband excitation. $\epsilon_{bg}$ represents a background dielectric constant from the metal's core-shell electrons. Figure \ref{epsAg} shows the experimentally determined dielectric permittivity of a 100 nm thick silver foil on a glass microscope slide substrate.

	The amorphous silica layers are modelled using Sellmeier's equation with an Urbach absorption tail. The real part of the refractive index is defined with Sellmeier's equation while the extinction coefficient is treated with a exponential absorption tail with an onset $\approx 330$ nm. It is important to note that this model does not satisfy the Kramers-Kronig relation but does acurrately model the dielectric function of amorphous silica in the near-UV to infrared range. A more complete, physically reasonable description of the optical response of silica can be achieved with a Tauc-Lorentz model; see page 170 of \cite{Fujiwara2007}.

	\subsubsection{Transmission Measurements}
	Optical and UV transmission measurements were taken using a {\tt Perkin-Elmer Lambda 900 IR-UV} spectrophotometer equipped with a UV polarizer ($\lambda = 300-400$ nm) and an IR polarizer ($\lambda =400-800$ nm). Baseline measurements indicate that the precision of transmission measurements is $\pm0.2 - \pm0.4\%$.
	The precision of the Spectrophotometer is minimized by using longer detector intergration times and by allowing the source lamp to warm-up and stabilize for $\approx 30-60$ min.

\subsection{Lossy Part of the Dielectric Tensor for silver/silica superlattices}

This section shows the real and imaginary (lossy) components of the effective medium, metamaterial dielectric permittivity tensor for the three silver/silica ultilayer samples in our experiment.
The dielectric permittivity tensor is computed using zeroth order Maxwell-Garnett effective medium theory and yields
$\bar{\bar{\epsilon}}_{eff} = \textrm{diag}[\epsilon_{||},\epsilon_{||},\epsilon_\perp]$ where $\epsilon_{||} = \rho\epsilon_{m} + (1-\rho)\epsilon_{d}$ is the dielectric permittivity for polarizations along the interfaces of the multilayer and $\epsilon_\perp = \left(\rho/\epsilon_{m}+(1-\rho)/\epsilon_{d}\right)^{-1}$ is the dielectric permittivity for polarizations perpendicular to the interfaces of the multilayer.
	$\epsilon_m$ is the silver permittivity and $\epsilon_d$ is the permittivity of the silica. $\rho$ is the metal volume filling fraction. 

\begin{figure}[hb]
\centering
	\includegraphics[width=\columnwidth]{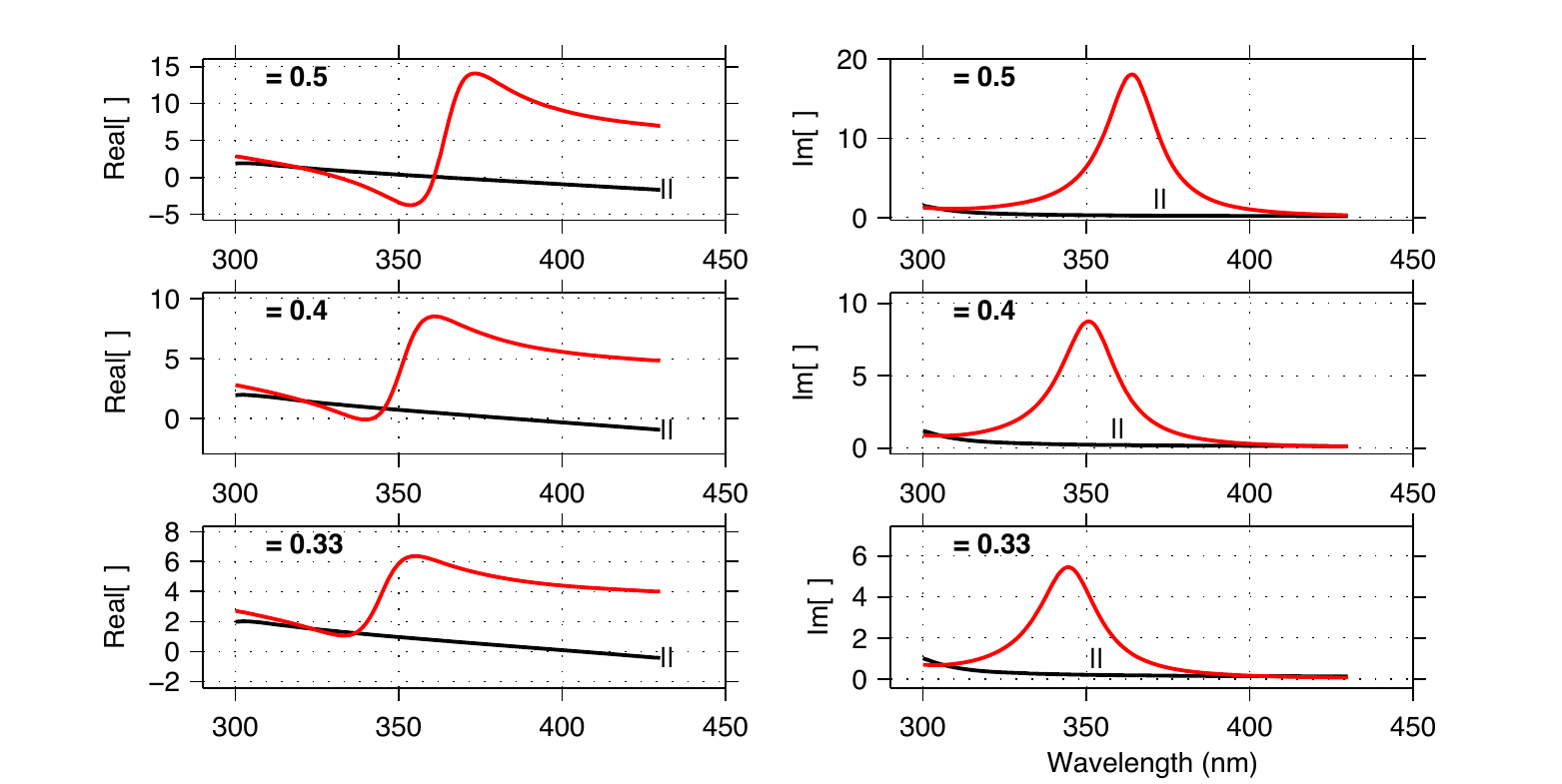}
	\caption{The real and imaginary part of the dielectric permittivity tensor for silver/silica superlattices is shown for various silver volume filling fractions. The ENZ in the parallel direction is red-shifted with decreasing silver filling fraction while the ENP in the perpendicular direction is blue-shifted with decreasing silver filling fraction. The dielectric tensor shows no remarkable features in the real or imaginary parts at the ENZ of the consituent silver ($\lambda\approx326$ nm).}
	\label{epsEMT}
\end{figure}

Figure \ref{epsEMT} shows the full metamaterial dielectric permittivity tensor for multilayer samples with silver film thicknesses of 20 nm and silica film thicknesses of 20 nm ($\rho=0.5$), 30 nm ($\rho=0.4$), and 40 nm ($\rho=0.33$).
The individual permittivity tensor components of the silver/silica multilayer-based metamaterials exhibit nothing remarkable at the ENZ of the constituent silver films which occurs at a wavelength of $\lambda = 326$ nm.
This reinforces the fact that complex propagation constant $k_z$ reveals underlying modal resonances of the metamaterial which can not be ascertained by examining the effective permittivty tensor alone.

\subsection{Modal Analysis - Finding the poles of $r_p$}

A mode supported by a particular structure can be characterized by a unique spectral frequency $\omega$ (or free space wavelength $\lambda$) and a spatial wavevector $\vec{k}$.
These modes are manifested as poles of the structure's response function, which for planar structures is the reflection (or transmission) coefficient.
The reflection and transmission coefficients have poles at the same locations and are thus equal candidates for finding the modes.
For a fixed wavelength $\lambda$ the reflection coefficient of a planar structure can be completely specified by the transverse wavevector $k_x$.
The energy-momentum (energy-wave vector) dispersion of a mode is then described by a path in the frequency-wavevector space, and at every point along this path, the reflection coefficient has a pole or equivalently, the inverse of the reflection coefficient has a zero. 

To determine the dispersion of modes supported by planar structures, we seek all the zeros (local minima) of the function $\alpha(k_x) = \left|1/r_p(k_x)\right|$ in the complex plane of $k_x = \beta + i\kappa$, and perform this search iteratively at a series of wavelengths $\lambda$.
An optimized transfer matrix method is used to calculate the reflection coefficient, while the minimum of the function $\alpha(k_x)$ is determined with the built-in MATLAB function \texttt{fminsearch}.
This built-in MATLAB function utilizes an optimized local-minimum search algorithm known as the Nelder-Mead simplex direct search method.
The algorithm is considered to have converged and the ``pole'' is found when the function value $\alpha(k_x)<10^{-4}$.

\begin{figure}
\centering
	\includegraphics[width=0.6\columnwidth]{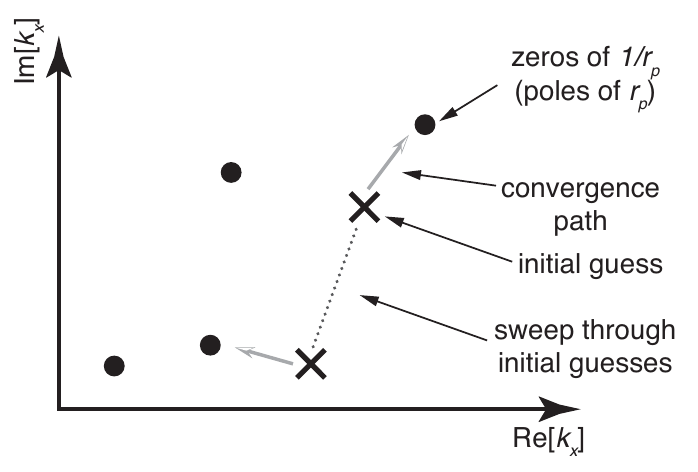}
	\caption{At a fixed wavelength, a structure can support many different modes characterized by a set of poles of the reflection coefficient $r_p$. These poles lie in the complex wavevector plane $k_x$. We search for the zeros of the function $\alpha=1/r_p(k_x)$ by systematically sweeping the initial guesses which are fed to the minimization algorithm \texttt{fminsearch} in MATLAB. The algorithm can converge on a different zero of $\alpha=1/r_p(k_x)$, depending on the initial guess. The black points are the location of the poles in the complex $k_x$-plane. The crosses are the initial guesses fed to the algorithm. The grey paths illustrate the pole to which the algorithm converges for a given initial guess.}
	\label{algorithm}
\end{figure}

Figure \ref{algorithm} illustrates the method used to determine all modes supported by a planar structure at a fixed wavelength.
Depending on the location of the initial guess in the complex wavevector plane that is fed to the algorithm, the algorithm will converge on different poles. 
We then interatively sweep through the initial guesses to ensure convergence of all poles (modes) at a given wavelength.
Physically reasonable ``leaky modes'' correspond to a complex transverse wavevectors $k_x = \beta + i\kappa$ such that the complex propagation constant $k_z = \sqrt{k_o^2-k_x^2}$ in the vacuum superstrate has Re$[k_z]>0$ and Re$[k_z] +$ Im$[k_z] > 0$.

\subsection{Ferrell-Berreman mode in thin- vs. thick-film}

\begin{figure}
	\includegraphics[width=0.6\columnwidth]{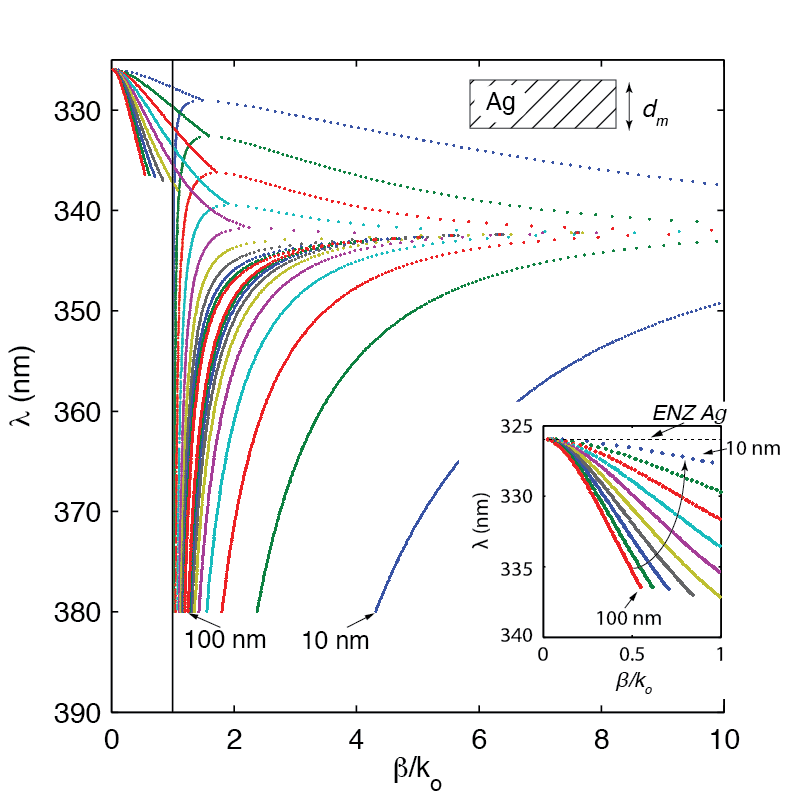}
	\caption{The evolution of the Ferrell-Berreman mode, the long-range SPP, and the short-range SPP for a thick silver foil to a thin silver foil. For thick films ($d_m\geq100$ nm) the long-range SPP and short-range SPP are degenerate and as the foil thickness decreases the dispersion of these two modes ``split'' into disctintly different modes. For very thin films of silver, the Ferrell-Berreman mode has a very slow, negative group velocity that couples with free-space. For thicker films, the Ferrell-Berreman mode has a much larger negative group velocity and couples poorly with free-space light.}
	\label{evolution}
\end{figure}

Figure \ref{evolution} shows the dispersion of the Ferrell-Berreman mode along with the dispersion of the long- and short-range surface plasmon polaritons that are supported by various thickness silver films.
When the silver foil thickness is greater than a few skin depths of silver ($d_m\geq100$ nm), the film supports two independent surface plasmons on either side of the foil which weakly interact with each other (see ``100 nm'').
As the film thickness decreases these modes interact and split into two distinct surface plasmons: the long- and short-range surface plasmons (see ``10 nm'').
For thick-films the Ferrell-Berreman mode (see inset) has a large negative group velocity, large imaginary part of its transverse wavevector Im$[k_x]$ and this mode couples poorly with free-space light.
As the film thickness decrease, the Ferrell-Berreman modes becomes a slow-light mode and the imaginary part of its transverse wavevector Im$[k_x]$ is significantly reduced, thus leading to the Ferrell-Berreman mode interacting strongly with free-space light in the case of a very thin silver foil.
In the effective medium, metamaterial limit the silver foils making up the metamaterial are considered to have vanishingly thin thicknesses.

\end{document}